\documentclass[reprint,prd,twocolumn,superscriptaddress,showpacs,nofootinbib,floatfix,preprintnumbers]{revtex4-1}
\usepackage{graphicx,bm}
\usepackage{color}
\usepackage{hyperref} 
\hypersetup{linktocpage=true}
\usepackage[all]{hypcap}
\hypersetup{bookmarks=true,         
   unicode=false,          
   pdftoolbar=true,        
   pdfmenubar=true,        
   pdffitwindow=false,     
   pdfstartview={FitH},    
   pdftitle={My title},    
   pdfauthor={Author},     
   pdfsubject={Subject},   
   pdfcreator={Creator},   
   pdfproducer={Producer}, 
   pdfkeywords={keyword1} {key2} {key3}, 
   pdfnewwindow=true,      
   colorlinks=true,       
   linkcolor=blue,          
   citecolor=magenta,        
   filecolor=magenta,      
   urlcolor=cyan           
}

\begin{document}
\preprint{DAMTP-2014-52} 
\title{Interpreting a CMS $lljj \slash\!\!\!p_T$ Excess With the Golden
  Cascade of the MSSM}

\author{Ben Allanach}
\affiliation{DAMTP, CMS, Wilberforce Road, University of Cambridge, Cambridge, CB3 0WA, United Kingdom}
\author{Anders Kvellestad}
\affiliation{Department of Physics, University of Oslo, N-0316 Oslo, Norway} 
\author{Are Raklev}
\affiliation{DAMTP, CMS, Wilberforce Road, University of Cambridge, Cambridge, CB3 0WA, United Kingdom}
\affiliation{Department of Physics, University of Oslo, N-0316 Oslo, Norway} 

\begin{abstract}
The CMS experiment recently reported an excess consistent with an invariant mass
edge in 
opposite-sign same flavor  (OSSF) leptons, when produced in conjunction
with at least two jets and missing transverse momentum. 
We provide an interpretation of the edge in terms of (anti-)squark pair
production followed by the `golden cascade' decay for one of the squarks:
$\tilde q 
\rightarrow 
\tilde\chi_2^0 q \to \tilde l l q \rightarrow  \tilde\chi_1^0 q
l l $ in the minimal supersymmetric standard model (MSSM).  
A simplified model involving binos, winos, an on-shell slepton, and the first
two generations of 
squarks fits the event rate and the invariant mass edge.
We check consistency with a recent ATLAS search in a similar region, finding
that much of the good-fit parameter space is still allowed at the 95$\%$
confidence level (CL). However, a combination of other LHC searches, notably
two-lepton stop pair searches and jets plus $\slash\! \! \! p_T $, rule out all of
the remaining parameter space at the 95$\%$ CL.
\end{abstract}
\maketitle

\section{Introduction}
A recent CMS search for beyond the standard model physics in a channel with at least two leptons, at least two jets
and missing transverse momentum ($\slash \!\!\! p_T$), reports a
2.6$\sigma$ excess~\cite{CMS-PAS-SUS-12-019} for 19.4 fb$^{-1}$ of integrated
luminosity at a centre of mass energy of 8 TeV.\footnote{After the initial completion of this paper the CMS experiment published a more complete account of the results of  that search in~\cite{Khachatryan:2015lwa}.} The signal consists of two 
isolated OSSF
leptons $l$ ($e$ or $\mu$). 
$e \mu$ opposite sign opposite-flavor (OSOF) leptons
are used to measure the backgrounds accurately. These are dominated by $t \bar
t$ production, which 
gives equal rates for the same-flavor and opposite-flavor
channels. Drell-Yan production of $\gamma^*$/$Z^0$ bosons is a secondary
irreducible background, yielding same-flavor events, and is estimated by
a control region in the event kinematics which does not overlap with the
signal region. The ATLAS experiment has looked in a similar signal region as CMS~\cite{Aad:2015wqa}, and seen no
excess, and so the two experimental results appear at first sight to be in
tension with one 
another.

\begin{figure}[t]
  \includegraphics[width=0.5 \textwidth]{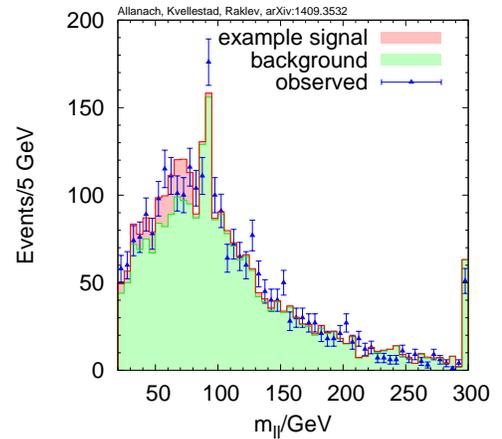}
  \vspace{-0.5cm}
  \caption{Invariant mass distribution of OSSF leptons in the 
    CMS selection after cuts. The expected Standard Model background is shown (green), 
    which is calculated from data by using OSOF events, 
    as well as
    the observed data and an example signal point (red)  in the parameter space
    investigated here involving the golden cascade: $m_{\tilde
      q}=900$ GeV, $m_{\tilde\chi_2^0}=312$ GeV, $m_{{\tilde l}_R}=200$ GeV, 
    $m_{\tilde\chi_1^0}=216$ GeV. Error bars
    on the observed number of events show the expected statistical standard
    deviation.} \label{fig:mll}
\end{figure}

\begin{figure}
  \includegraphics[width=160pt]{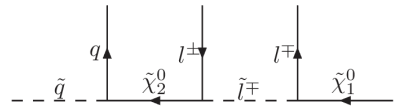}
  \vspace{-0.5cm}
  \caption{Feynman diagram for the golden cascade decay.}
  \label{fig:golden}
\end{figure}

The CMS excess over the Standard Model expectation is depicted in
Fig.~\ref{fig:mll} and shows an interesting kinematical feature: the invariant
mass of the lepton pair $m_{ll}$ is consistent with a right triangular shaped
kinematic edge at $m_{ll}=78.7 \pm 1.4$~GeV~\cite{CMS-PAS-SUS-12-019}.
Features such as edges are less likely to come 
from mis-modelling the detector response to backgrounds than smoother shapes,
and so they are particularly welcome as indicators of a signal. 
This triangular edge is a classic signal of the production of supersymmetric
(SUSY) particles 
which undergo two-body cascade decays through successively lighter
on-shell SUSY
particles, for example the chain $\tilde\chi_2^0 \rightarrow {\tilde l}^- l^+ \rightarrow
\tilde\chi_1^0 l^-l^+$. 
The jets in the signal events could either be the result of initial state radiation, or of the
$\tilde\chi_2^0$ being produced itself by the decay of  a squark $\tilde q \rightarrow \tilde\chi_2^0 q$. This golden chain,
starting from the squark, see Fig.~\ref{fig:golden}, has been intensely studied
for the possibilities it brings for determining the parameters of the
sparticles involved, such as mass and spin. 
For a review see~\cite{Barr:2010zj}. 

The MSSM predicts that the LHC produces 
pairs of SUSY particles, {\it e.g.}\ squarks and neutralinos, each
with various possible decay chains. As an interpretation of the excess CMS gave three
benchmark model points with a sbottom squark in the cascade 
decay chain~\cite{CMS-PAS-SUS-12-019}.\footnote{In this chain,
the $\tilde\chi_2^0$ decays through an off-shell $Z$, which does not
predict an exact triangular   di-lepton 
  distribution~\cite{Miller:2005zp,Lester:2006cf}.}   
They showed that the predicted 
$m_{ll}^{max}$ distribution was roughly in agreement with data for two of
their benchmarks but provided no scan of the parameter space or other tests of the benchmarks.  Production from sbottoms was investigated further in~\cite{Huang:2014oza}. 

Here, we shall instead interpret the
excess in terms of the production of
squarks from the first two generations, and provide a more comprehensive
exploration of the interesting parameter space in Sec.~\ref{sec:pspace}.
The null results of the corresponding ATLAS search and strong direct constraints on light flavoured squarks from
LHC searches for jets and 
$\slash \!\!\! p_T$ and no leptons, will have an impact on the allowed parameter space. In Sec.~\ref{sec:constraints} we shall investigate whether the interpretation of this CMS excess
involving the golden channel is consistent with these and other collider
constraints. Finally, we draw our conclusions in Sec.~\ref{sec:conclusion}.

\section{Parameter space fitting the CMS excess}
\label{sec:pspace}
The edge in $m_{ll}$ predicted by the $\tilde\chi_2^0$ decay chain is due to
kinematics: one finds~\cite{Allanach:2000kt}, by 
energy-momentum conservation, that in the decay chain described
above with an on-shell slepton $\tilde l$, it has a 
maximum value 
\begin{equation}
m_{ll}^{max} = \sqrt{\frac{(m_{\tilde\chi_2^0}^2 - m_{\tilde l}^2)(m_{\tilde
      l}^2 - m_{\tilde\chi_1^0}^2)}{m_{\tilde l}^2}}. \label{eq:edge}
\end{equation}
Thus, measurement of the edge leads to a constraint upon the masses of the
three 
SUSY particles involved in the decay.

\begin{figure}
  \includegraphics[width=8.0cm]{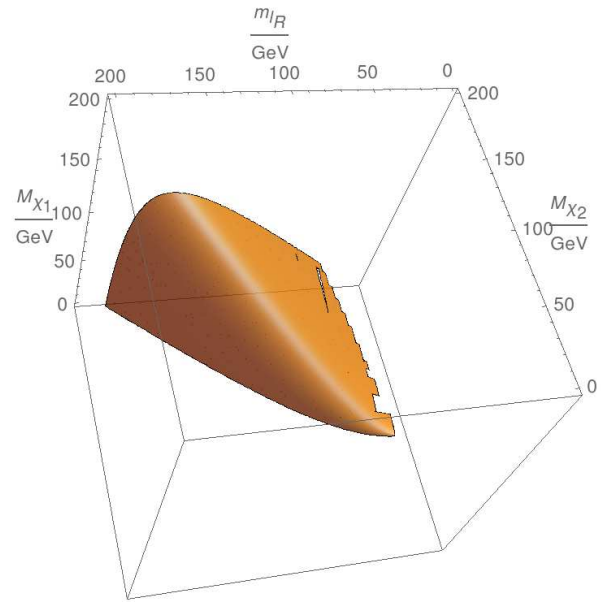}
  \caption{Constraint on SUSY particle masses involved in the
    $\tilde\chi_2^0$ decay coming from the central value $m_{ll}^{max}=78.7$
    GeV.} \label{fig:edge}
\end{figure}

We show the edge constraint on the masses coming from the central value
inferred from CMS data in Fig.~\ref{fig:edge}. 
From the endpoint constraint alone the  hyper-surface will
extend to infinite masses, while from below it only bounds the mass of $\tilde\chi_2^0$. The errors on the
CMS fit to the edge are so small that varying $m_{ll}^{max}$ within them would produce no visible
difference in the figure. 

In our interpretation we follow the CMS counting experiment analysis where two
OSSF leptons are required to have transverse momentum $p_T>20$ GeV and
pseudorapidity $|\eta|<2.4$, excluding the range $1.4 < |\eta| < 1.6$ where
electron and muon efficiencies differ greatly. Jets are reconstructed by
the anti-$k_T$ algorithm~\cite{Cacciari:2008gp} using {\tt
  FastJet}~\cite{Cacciari:2011ma}, with a jet radius parameter of $R=0.5$, and
are required to have $p_T>40$ GeV and lie within $|\eta|<3.0$. 
A combination of two jets and missing transverse momentum $\slash \!\!\!
p_T>150$ GeV, or three or more jets and $\slash \!\!\! p_T>100$ GeV, is
required in the events. For di-lepton invariant masses in the range 20 GeV $<
m_{ll}<70$ GeV the total CMS background estimate is 730$\pm 40$ events for
central 
production (both leptons within $|\eta|<1.4$), whereas 860 OSSF were observed,
corresponding 
to a 2.6$\sigma$ deviation. The deviation 
of $130^{+48}_{-49}$ events constrains the MSSM parameter
space.

For given $m_{\tilde\chi_1^0}$ and $m_{\tilde\chi_2^0}$ masses and the measured $m_{ll}^{max}$,
there are at most two possible positive real solutions of Eq.~(\ref{eq:edge})
for 
$m_{{\tilde 
    l}}$. In the 
rest of this work we shall pick $m_{{\tilde l}}$ and either $m_{\tilde
  \chi_1^0}$ or $m_{\tilde \chi_2^0}$ by
changing an input 
parameter, then impose 
Eq.~(\ref{eq:edge}) by solving it for the other neutralino mass. Then, the
overall interpreted signal {\em rate}\ gives the mass for the squarks: the
heavier they are, the smaller the production cross section and the smaller the
rate.  

We shall use a bottom-up prescription in order to fit the CMS excess, setting
MSSM particles that are irrelevant for the signal to be heavy. 
We use as free parameters the wino soft-mass $M_2$, a common
first and second generation\footnote{CMS did not release a flavor
  decomposition of the events. Given more statistics, this can be used
to infer a possible smuon-selectron mass splitting~\cite{Allanach:2008ib}.} 
right-handed soft mass $m_{\tilde l_R}$, solving
for the correct value of the bino soft-mass $M_1$,\footnote{We consider both
  hierarchies: $M_2>M_1$ (bino dominated LSP) and $M_2<M_1$ (wino dominated
  LSP). Higgsinos only couple extremely weakly to selectrons or smuons and so
  would result in rates that were far too small if they were involved in the
  chain.}  and a common 
first and second generation squark mass (both left- and right-handed)
$m_{\tilde q}$. The mass of the SUSY partner of the left-handed lepton
$m_{{\tilde l}_L}$ is fixed to be $2 m_{\tilde l_R}$.
Setting $m_{{\tilde l}_L} < m_{\tilde\chi_2^0}$ would introduce the $\tilde l_L$
into the decay chain, as well as light sneutrinos
that steal branching ratio from the golden cascade, and thus lower the
signal rate. 

\begin{figure}
  \includegraphics[width=8.0cm]{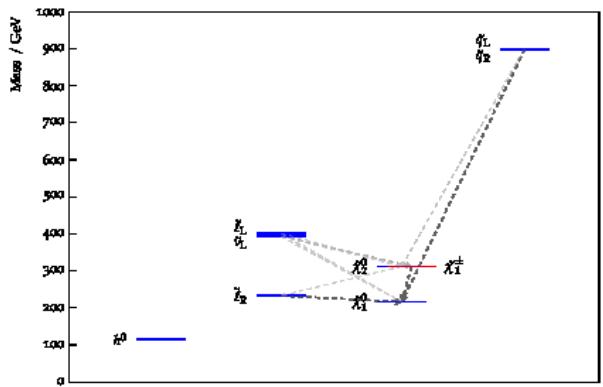}
  \caption{Example signal point that fits the edge
    inference: $m_{\tilde
      q}=900$ GeV, $m_{\tilde\chi_2^0}=312$ GeV, $m_{{\tilde l}_R}=200$ GeV, 
    $m_{\tilde\chi_1^0}=216$ GeV.
    Prominent decays with branching ratios higher than $10\%$ are shown
    as arrows.}    
  \label{fig:spec}
\end{figure}

Except the gluino mass, which is set to 1.6 TeV, all other soft masses are decoupled at\footnote{Instead of fixing the Lagrangian parameters for the soft SUSY
  breaking Higgs mass parameters, we calculate them by minimising the MSSM
  Higgs potential after fixing $M_A$ and the $\mu$ parameter to 3500
  GeV~\cite{Skands:2003cj}.} 3500
GeV, and the 
trilinear soft SUSY breaking scalar couplings are set to zero. 
Decoupling the gluino mass makes it easier for the scenario to
pass constraints from searches in the jets plus $\slash \!\!\!p_T$ channel,
however, an alternative interpretation could potentially be found by
decoupling the squarks instead. 
We also set $\tan\beta=10$. Although this is a parameter in
the neutralino mass matrix we have checked that changing $\tan\beta$ has
a negligible effect on our CMS fit. We show an example spectrum, along with
prominent decays, in Fig.~\ref{fig:spec}.
 
We calculate the resulting sparticle spectrum using {\tt SOFTSUSY 3.5.1}~\cite{Allanach:2001kg} and the sparticle branching ratios with {\tt  SUSYHIT 1.4}~\cite{Djouadi:2006bz}. Spectrum
and decay information is communicated via the SUSY Les Houches
Accord~\cite{Skands:2003cj}. For given values of $m_{{\tilde l}_R}$ and $M_2$, $M_1$ is calculated to solve
  Eq.~(\ref{eq:edge}). 

\begin{figure}
  \includegraphics[width=8.0cm]{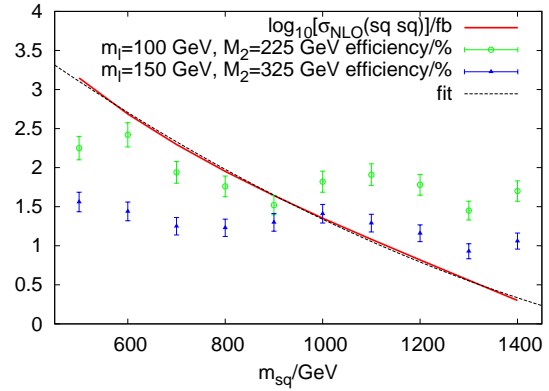}
  \caption{Fit to (anti-)squark production cross-section at an 8 TeV LHC\@. The
  logarithm of the {\tt Prospino} cross-section prediction is shown by the
  solid line. Our quadratic fit to this is shown by the dashed
  line. The
  CMS $lljj \slash \!\!\! p_T$ search region cut efficiencies
  for example parameter space points are shown by the
  points with errorbars, where the errorbars are purely from Monte Carlo
  statistics.}     
  \label{fig:fitsig}
\end{figure}

We calculate the production cross-section of squarks and anti-squarks to
next-to-leading order for these parameter points using a fit to results from {\tt Prospino}~\cite{Beenakker:1996ch}.  Figure~\ref{fig:fitsig} shows the prediction for the production cross-section of
(anti-)squarks at an 8 TeV LHC\@. We have fitted a function such that
\begin{equation}
\log_{10} \sigma /\textrm{fb} = 
a_2 x^2 + a_1 x + a_0, \label{quad}
\end{equation}
where $x=m_{\tilde q}(M_{SUSY}) / (\textrm {1 TeV})$ is proportional to the
squark mass input 
parameter. The squark mass input depends upon the modified $\overline{DR}$ mass 
renormalisation scale, which we have set equal to the geometric mean of the
two stop masses ($M_{SUSY}$).
Our fit yields $a_2=1.12855$, $a_1=-5.22317$ and $a_0=5.43447$.
The production cross-section varies rapidly with squark mass: in the mass
region checked, a factor of 3 in squark mass results in 3 orders of magnitude
reduction in the (anti-)squark production cross-section.

We then generate 40\,000 SUSY Monte Carlo events
per parameter point using {\tt Pythia 8.186}~\cite{Sjostrand:2006za,Sjostrand:2007gs}. 
These events are
propagated through our implementation of the CMS analysis.
Figure~\ref{fig:fitsig} shows that the efficiency of the cuts 
of the CMS $lljj \slash\!\!\! p_T$ signal region (see below) varies much more
slowly with squark mass input parameter than the cross section. This suggests a strategy for finding
the correct squark mass to yield a desired signal yield: we first calculate
the number of expected events in the CMS signal region for an initial input
squark mass parameter (we take 1000 GeV). Then we calculate $\sigma$ needed
for our desired signal yield, assuming that the efficiency does not change.  
We solve Eq.~(\ref{quad}) for $x$, set the squark mass input parameter to 
$x \times 1$ TeV, then calculate the cut efficiency from the sample of
simulated signal 
events. This process is iterated until the squark mass converges, allowing us
to efficiently find points in the three-dimensional $m_{{\tilde l}_R}$,
$\Delta m=M_2 - m_{{\tilde l}_R}$,
$m_{\tilde q}(M_{SUSY})$ input parameter space that correspond to a given
number of CMS $lljj \slash\!\!\! p_T$ signal region events. 
Convergence here is defined as the input squark mass changing by less than 10
GeV between the previous iteration and the present.

The predicted $m_{ll}$ distribution of an example point that fits the
inferred edge along with the rate is shown in Fig.~\ref{fig:mll}. It
is consistent with the CMS $lljj \slash \!\!\! p_T$ data. 
However, we shall eventually show that a combination of constraints will rule
out the 
golden channel interpretation of the excess to 95$\%$ CL\@. In order to make
this interpretation robust, we wish to show that even CMS $lljj\slash \!\!\!
p_T$ signal rates at the 95$\%$ lower boundary are excluded: higher signal
rates would result from higher production cross-sections, {\it i.e.}\ lower squark
masses, but this would then produce higher rates for the other searches,
disfavouring the golden channel interpretation even more. Profiling over
Gaussian background uncertainties, the observation of 860 events in the signal
region over a background of $730 \pm 40$ yields a 95$\%$ CL lower limit of 34
signal events. We will therefore find the parameter space
corresponding to this number of predicted signal events. 

\begin{figure}[t]
   \includegraphics[width=0.5\textwidth]{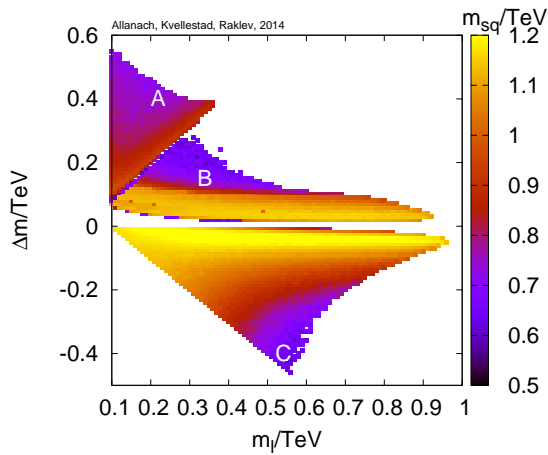}
   \caption{Constraints on golden channel parameter space from the CMS $lljj
     \slash \!\!\! p_T$ search over the
     input parameter plane $m_{\tilde l}(M_{SUSY})$ and $\Delta
     m=M_2(M_{SUSY}) - m_{\tilde 
       l}(M_{SUSY})$. 
     A coloured
     box indicates a point that fits the 
      95$\%$ CL lower limit of the inferred value of CMS $lljj \slash 
     \!\!\! p_T$
     signal rate and is also consistent with CMS's 
     edge inference.}
   \label{fig:pars}
 \end{figure}

In Fig.~\ref{fig:pars} we show the region of input
parameter space that fits the CMS $lljj\slash \!\!\!
p_T$ signal rates at the 95$\%$ lower boundary. Wherever a coloured point is
plotted,  
there is a viable solution. Blank regions of the plot either require squark
masses that are below 500 GeV in order to get high enough signal rates, or do
not contribute to the signal because decays do not give the required
topology. For $\Delta m < 0$, the LSP is wino-dominated, whereas for $\Delta
m>0$, it is
bino-dominated. We see that the two signs of $\Delta m$ are separated by
a region where the resulting leptons tend to be too soft to give appreciable
signal rates. 
The colour of the point gives the physical average light squark mass
by reference to the colour bar on the right-hand side. 
Squark masses up to 1200 GeV are predicted,
depending upon parameters. We have divided the parameter space up into three
connected regions: A, B and C as shown on the figure. We show the main decay
modes relevant to the golden channel for each region in Table~\ref{tab:modes}.

\begin{table}
\begin{tabular}{|c|l|} \hline
Region & mode \\ \hline
A & $\tilde q_L \rightarrow q \tilde\chi_2^0 \rightarrow q \tilde l_L l \rightarrow q l^+ l^- \tilde\chi_1^0$ \\
  & $\tilde q_R \rightarrow q \tilde\chi_1^0$ \\ \hline
B & $\tilde q_L \rightarrow q \tilde\chi_2^0 \rightarrow q \tilde l_R l \rightarrow q l^+ l^- \tilde\chi_1^0$ \\
  & $\tilde q_R \rightarrow q \tilde\chi_1^0$ \\ \hline
C & $\tilde q_R \rightarrow q \tilde\chi_2^0 \rightarrow q \tilde l_R l \rightarrow q l^+ l^- \tilde\chi_1^0$ \\
  & $\tilde q_L \rightarrow q \tilde\chi_1^0$ \\ 
\hline\end{tabular}
\caption{Main relevant decay modes of each connected region in
  Fig.~\protect\ref{fig:pars}. \label{tab:modes}}
\end{table}

The location of region A, and in particular the slope dividing it from region B, of course depends on the relationship $m_{{\tilde l}_L}=2 m_{\tilde l_R}$ that we have fixed. However, the exact value of that slope will have little effect on the following discussion as long as $m_{{\tilde l}_L}> m_{\tilde l_R}$.
In region C, where $\Delta m<0$, the LSP is wino-dominated and so there is a
sizeable branching fraction for squarks to decay via the lightest charginos instead.  

\begin{figure}[th]
   \includegraphics[width=0.5\textwidth]{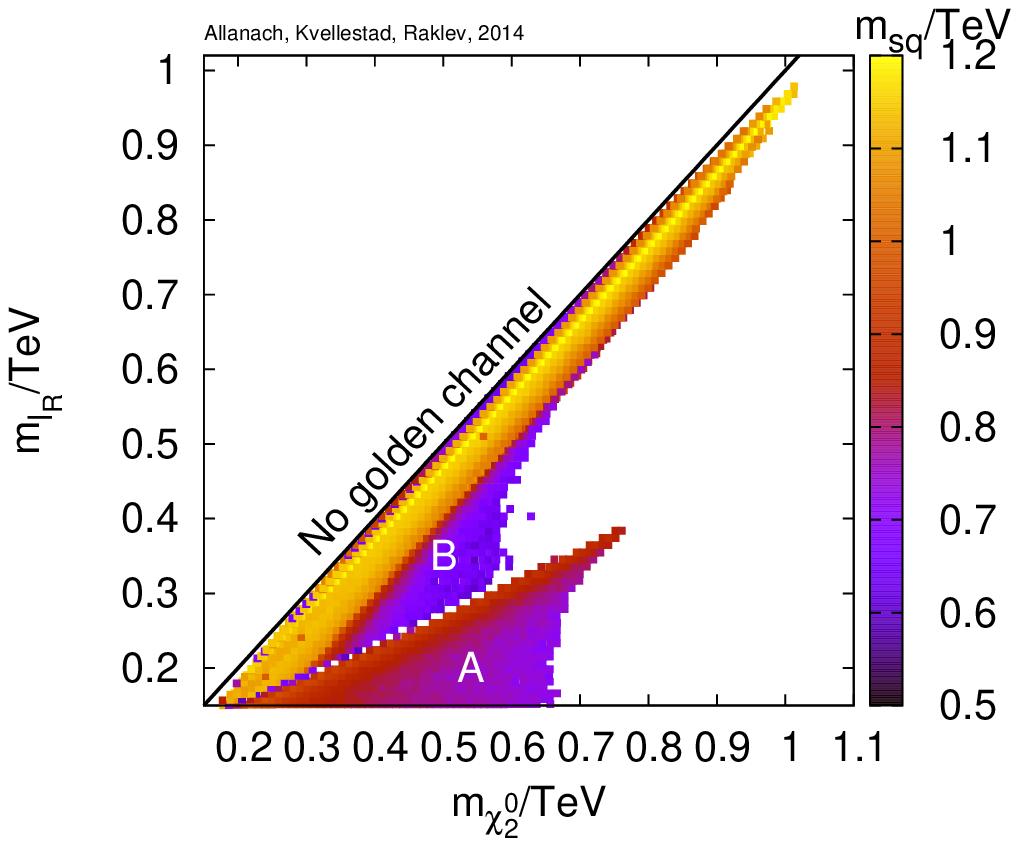}
   \includegraphics[width=0.5\textwidth]{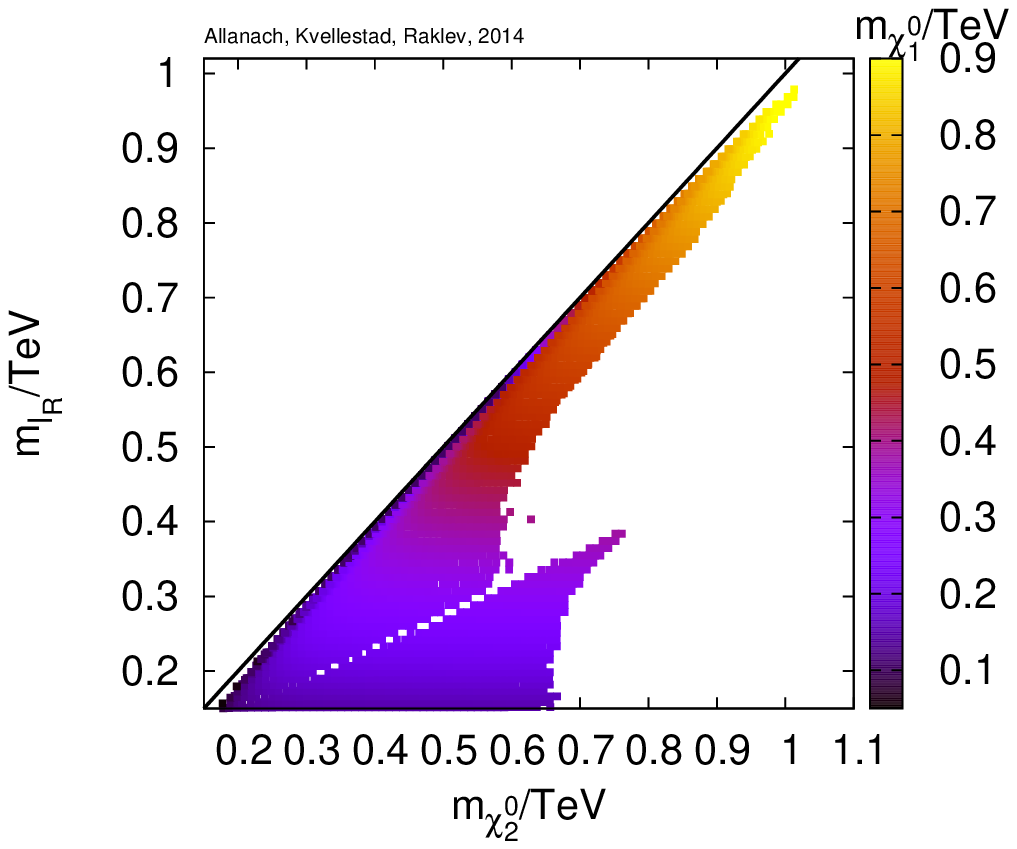}
   \caption{Constraints on golden channel parameter space from the CMS $lljj
     \slash \!\!\! p_T$ search over the
     plane of pole masses of the second-lightest neutralino and the
     right-handed slepton. 
     A coloured
     box indicates a point that fits the 
      95$\%$ CL lower limit of the inferred value of CMS $lljj \slash 
     \!\!\! p_T$
     signal rate and is also consistent with CMS's 
     edge inference. 
     The squark mass (upper panel) or lightest neutralino mass (lower panel)
     is given by reference to the scale on the  
     right-hand side. Above the black line, the golden channel is
     kinematically inaccessible.}
   \label{fig:edgeConst}
 \end{figure}

Figure~\ref{fig:edgeConst} displays 
the same points plotted as functions of the physical masses of the
second-lightest neutralinos and the right-handed sleptons. In the upper panel,
we display the physical squark mass that fits the 95$\%$ CL lower inferred
rate and in the lower panel, the lightest neutralino mass as inferred from the
edge constraint in Eq.~(\ref{eq:edge}).
There is an upper bound on
slepton masses 
$m_{{\tilde l}_R}<1000$ GeV implied by the fit. This is because in order to
get a sizeable decay rate for the golden cascade, we require the mass ordering
$m_{\tilde q}>m_{\tilde\chi_2^0}>m_{{\tilde l}_R}>m_{\tilde\chi_1^0}$ and for
such high $m_{\tilde q}$, 
it is no longer possible to get a large enough signal event rate. 
$m_{\tilde\chi_1^0}$ is highly correlated with $m_{{\tilde l}_R}$ in order to get
the central inferred $m_{ll}^{\max}$ value, and lies in the range 50 GeV to 800
GeV. In the upper panel, we label where the regions A, B or C are mapped to on
the physical mass plane. In fact, region C is mapped to a small region
close to the `No golden channel line' on top of region B in the physical mass
plane where $\tilde\chi_2^0$ and ${\tilde l}_R$ are virtually degenerate.

\section{Constraints from other searches}
\label{sec:constraints}
Both ATLAS~\cite{Aad:2014wea} and CMS~\cite{Chatrchyan:2014lfa}
have searched in the jets and $\slash \!\!\!p_T$
channel. Neither experiment observed a significant excess, and the 
exclusions from each are rather similar. Here, we constrain our parameter
space with an ATLAS search at 8 TeV in 20.3 fb$^{-1}$ of integrated
luminosity~\cite{Aad:2014wea} in the `3j' signal region. This signal region is chosen to be efficient for the type of events with a low number of high-$p_T$ jets expected from the topologies in Table~\ref{tab:modes}. 

Any events with
isolated muons
or electrons are vetoed, and ATLAS requires $\slash \!\!\!p_T>160$
GeV, and the three hardest anti-$k_T$ jets with $|\eta|<2.5$ and  $R=0.4$ to have at least 130 GeV, 60
GeV and 60 GeV, 
respectively. Their azimuthal angle must differ from that of the reconstructed 
$\slash \!\!\! p_T$ by $\Delta \phi > 0.4$. Defining the effective mass 
$m_{eff}$ as the scalar sum of $\slash \!\!\!p_T$ and the $p_T$ of the
hardest three jets, the cuts
 $\slash \!\!\!p_T/m_{eff}>0.3$ and $m_{eff}>2200$ are also imposed. ATLAS
 observed 7 events on a background of $5.0 \pm 1.2$, from which they deduce an
 upper bound of 8.2 signal events to 95$\%$ CL\@. We impose this constraint upon
 our expected signal yields, having checked that our implementation of the analysis is consistent with ATLAS results in terms of cut-flow. 
 
\begin{figure}[t]
   \includegraphics[width=0.5\textwidth]{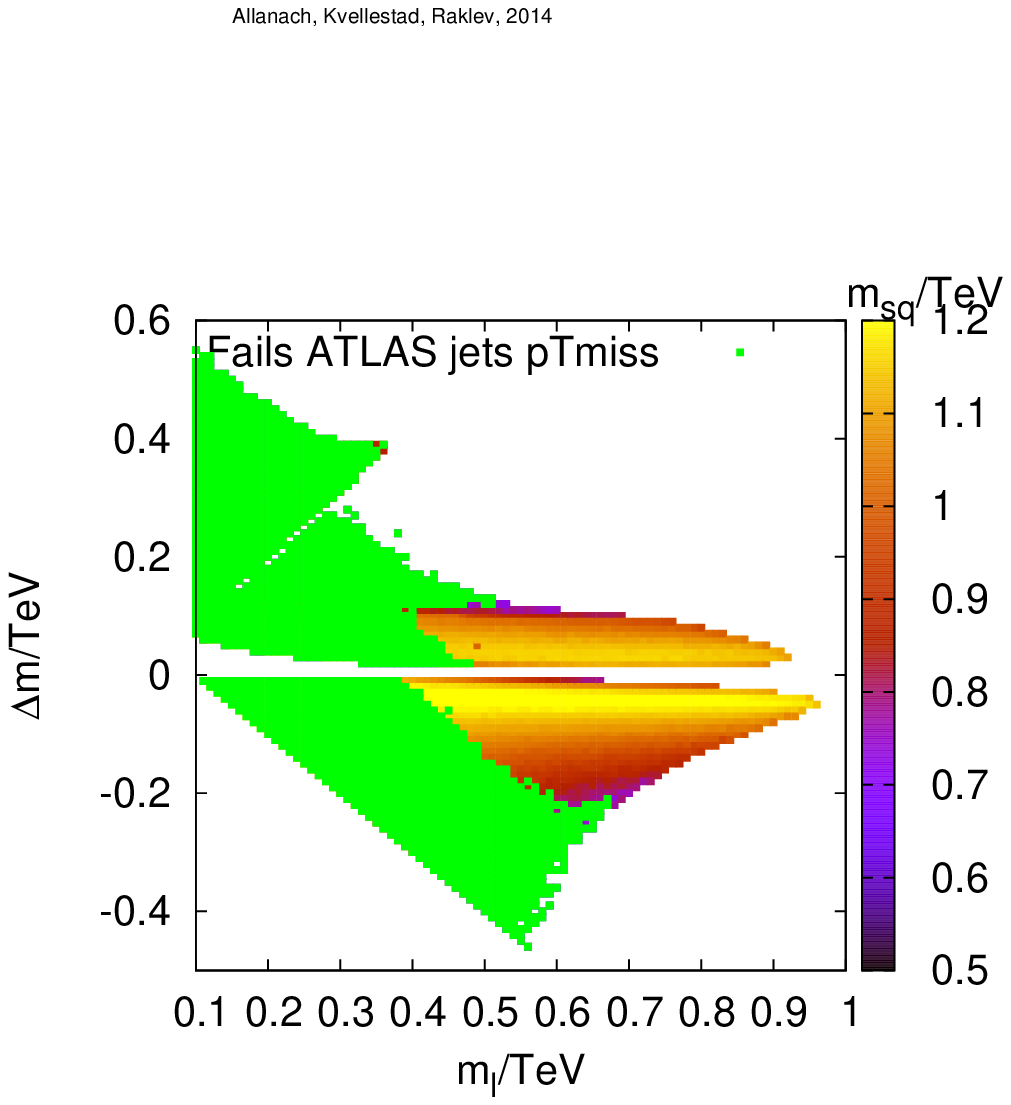}
   \caption{Constraints on golden channel parameter space from the ATLAS jets plus
     $\slash \!\!\! p_T$ search.
     The squark mass consistent with the 95$\%$ CL lower bound on 
     the CMS $lljj \slash \!\!\! p_T$ signal 
     is given by reference to the scale on the 
     right-hand side. Points coloured green are excluded by an ATLAS
     $\slash \!\!\! p_T$ plus jets search at the $95\%$
     CL\@. } 
   \label{fig:jmet}
 \end{figure}

Figure~\ref{fig:jmet} shows that 
a large fraction of otherwise viable parameter space is excluded by the
jets plus $\slash \!\!\! p_T$ constraint, but that a portion of parameter
space with $m_{\tilde l}>400$ GeV survives the constraint, despite having
squark masses as low as 750 GeV. The potency of the jets plus $\slash
\!\!\!p_T$ search is reduced by the large leptonic branching ratio in this
region of the plot, and lower signal rates due to the fact that we have set
the gluino mass to be rather high at 1.6 TeV.\footnote{Feynman diagrams with
gluinos in the $t-$channel contribute to di-squark production.}

Since CMS published its 2.6$\sigma$ excess, ATLAS has checked a similar signal
region, which they call the `off-$Z$' region, demanding two isolated same
flavor leptons
with $p_T>25, 20$ GeV, respectively, $20<m_{ll}<70$ GeV, and at least two jets
with $p_T>35$ GeV and pseudorapidity $|\eta|<2.5$~\cite{Aad:2015wqa}. For two
anti-$k_T$ jets of distance parameter $R=0.4$, ATLAS requires
$\slash \!\!\! p_T > 150$ GeV and for three or more, $\slash \!\!\! p_T > 100$
GeV. ATLAS observed 1133 events in this (`SR-loose same-flavor combined')
signal region on an estimated background of $1190 \pm 40 \pm 70$, where the first quoted
uncertainty is statistical and the second is systematic. Combining the two
uncertainties in quadrature and profiling over an assumed Gaussian background
expectation, we derive a 95$\%$ CL upper limit on the number of signal events
in this signal region of 125.0. 

\begin{figure}[t]
   \includegraphics[width=0.5\textwidth]{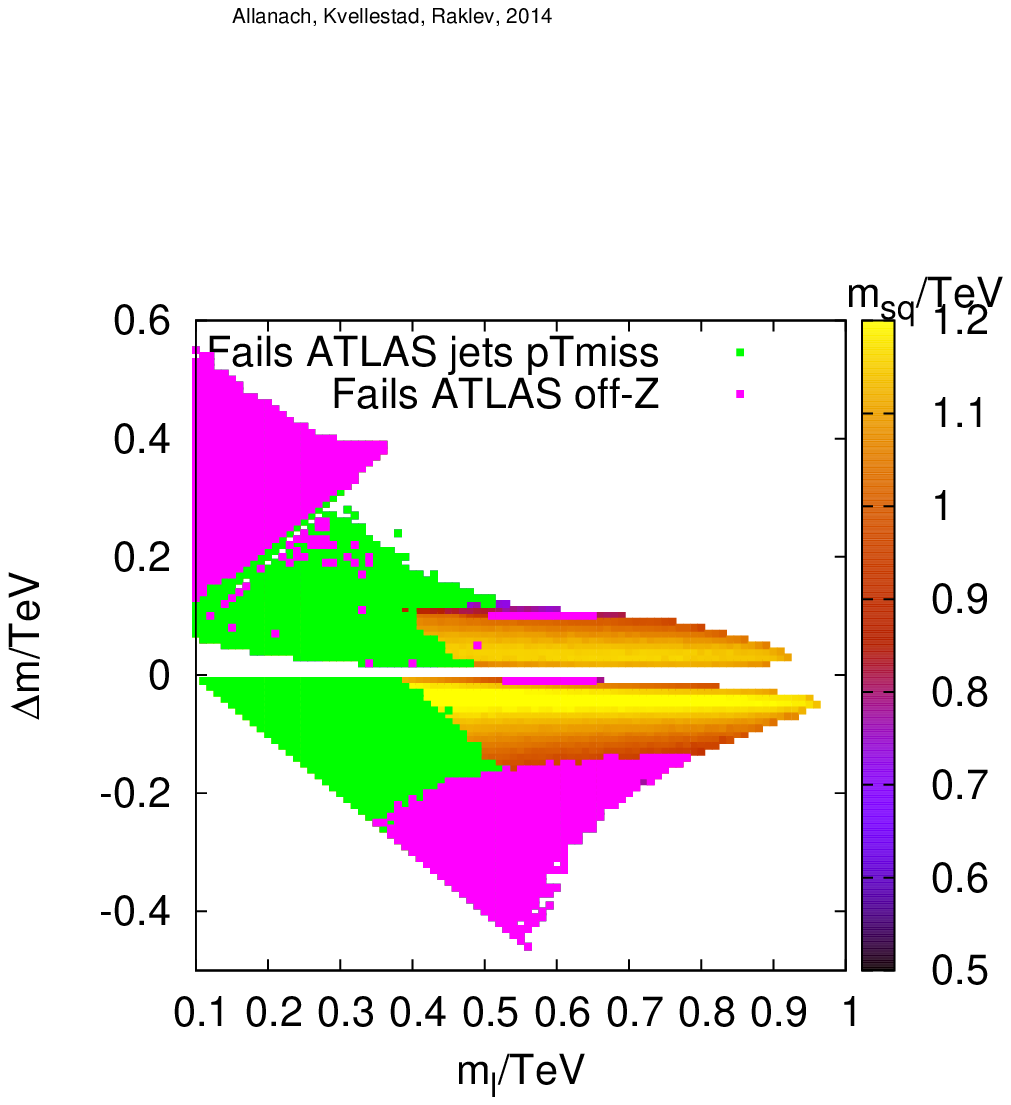}
   \caption{Constraints on golden channel parameter space from the ATLAS $lljj
     \slash \!\!\! p_T$ on-$Z$ search. 
     The squark mass consistent with the 95$\%$ CL lower bound on 
     the CMS $lljj \slash \!\!\! p_T$ signal 
     is given by reference to the scale on the 
     right-hand side. Magenta points are ruled out by the ATLAS search at the
     95$\%$ CL, whereas those coloured green  are ruled out by ATLAS
     $p_T\!\!\!\slash$ plus jets searches at the $95\%$
     CL\@.} 
   \label{fig:atlas}
 \end{figure}

Because the ATLAS cuts are slightly different to
those of  CMS, we must perform simulations in order to determine the ATLAS cut
efficiencies and see whether the upper limit on the number of signal events
constrains the parameter space significantly. Again our implementation of the analysis has been validated against ATLAS results. Figure~\ref{fig:atlas} shows that
the ATLAS search does constrain the part of golden-channel parameter space
that fits the CMS $lljj \slash \!\!\! p_T$ analysis to 95$\%$ CL, but that
there is still plenty of viable parameter space left. Most of the viable
parameter space ruled out by the ATLAS off-$Z$ search is also already ruled out by the jets
plus $\slash \!\!\! p_T$ search. 

Since we have chosen the parameters of our signal model to yield high
branching ratios of squarks to di-leptons plus $\slash \!\!\! p_T$, there is
the possibility of both squarks decaying via the di-leptonic cascade. This
then may predict a non-zero signal rate for four-lepton $\slash \!\!\!
p_T$ channels, which must be checked against experimental
searches. CMS~\cite{Chatrchyan:2014aea} placed bounds upon such channels by
requiring  
at least two OSSF
lepton pairs, $\slash \!\!\! p_T>100$ GeV,
and that neither pair is likely to come from a $Z-$boson, {\it i.e.}\ neither has
$75<m_{ll}/\textrm{GeV}<105$. The sample is split into a high 
energy region where the total
scalar sum of visible transverse momenta, $H_T>200$ GeV and a low energy
region where $H_T<200$ GeV.

\begin{figure}[t]
   \includegraphics[width=0.5\textwidth]{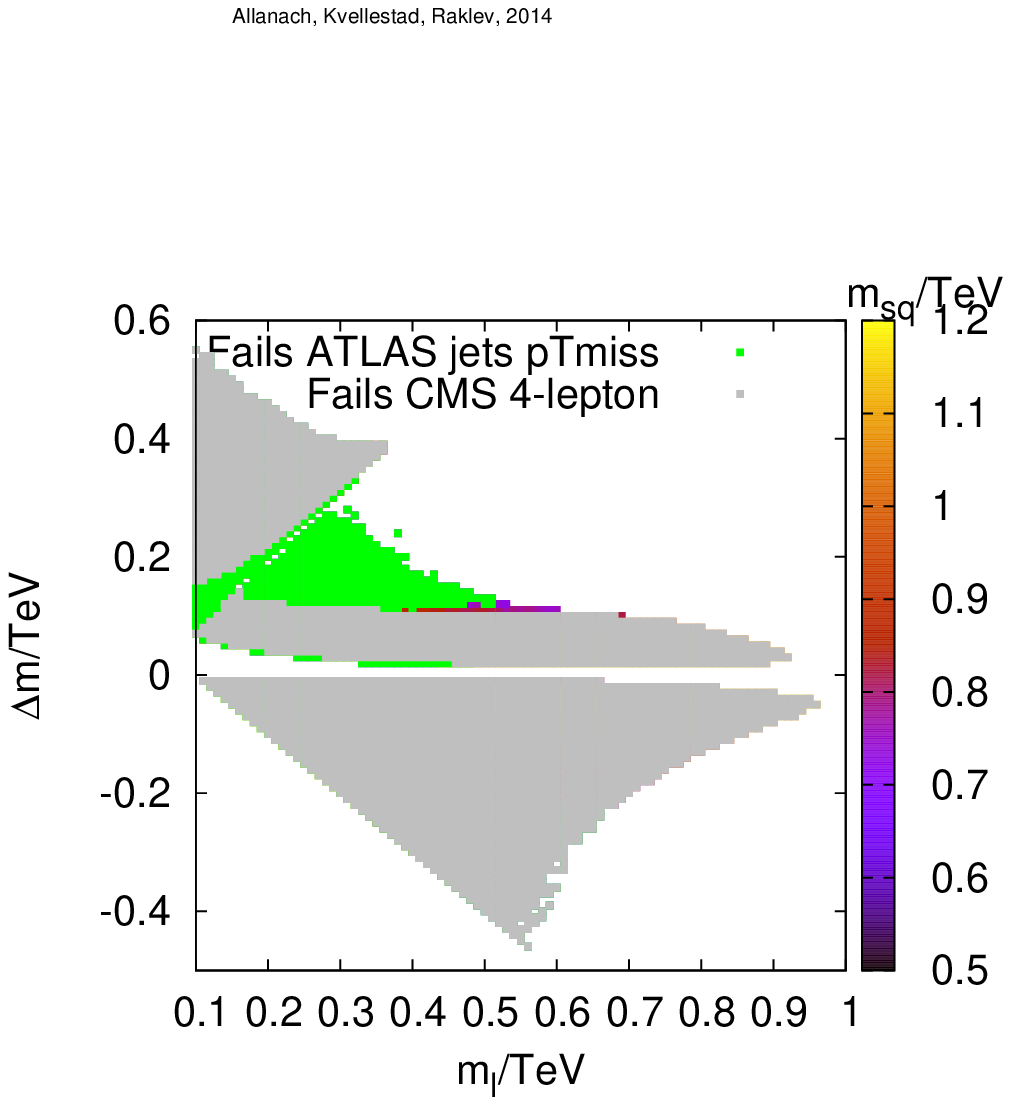}
   \caption{Constraints on golden channel parameter space from the CMS $4l
     \slash \!\!\! p_T$ search. 
     The squark mass consistent with the 95$\%$ CL lower bound on 
     the CMS $lljj \slash \!\!\! p_T$ signal 
     is given by reference to the scale on the 
     right-hand side. Light gray points are ruled out by the CMS 4-lepton
     search at the 
     95$\%$ CL, whereas those coloured green  are ruled out by ATLAS
     jets $\slash \!\!\! p_T$ searches, at the $95\%$
     CL\@.} 
   \label{fig:4l}
 \end{figure}

The most constraining signal
region expected for our hypothesised signal is the high energy region with
zero $b-$ or $\tau-$tags in addition.
In the high energy region, CMS observed zero events on a SM background expectation
of $0.01\pm 0.01$. We deduce a 95$\%$ CL upper bound on a
putative signal contribution of 3.0.
If a model point predicts an expected signal rate of larger than 3.0
events, we consider the point to be ruled out by these
four-lepton searches. 
The resulting constraints on the viable parameter space are shown in
Fig.~\ref{fig:4l}. We see that the four-lepton search places strong constraints
upon the model, ruling out nearly all of the remaining parameter space of the
model 
except for a thin sliver at $\Delta m \approx 120$ GeV and
$0.4<m_{\tilde l}/\textrm{TeV}<0.6$. This small remaining sliver is where the
branching ratio of the golden channel decay 
$BR({\tilde q}_L \rightarrow \chi_2^0 q \rightarrow {\tilde l^\pm}_R l^\mp q \rightarrow
\chi_1^0 l^+ l^-q)$ is less than around 6$\%$, resulting in lower four-lepton
signal rates. 

The most relevant other search to our golden channel interpretation of the CMS
excess is one by ATLAS for direct stop pair production in final states with
two leptons.\footnote{The strong effect of this search was first pointed out
  by the authors of~\cite{Grothaus:2015yha}. We include it here for
  completeness following a revision of our original preprint.} ATLAS searched
in 20.3 fb$^{-1}$ of integrated luminosity of 8 
TeV $pp$ collisions, in channels with exactly two oppositely
charged leptons with $p_T>25, 20$ GeV, respectively, and $m_{ll}>20$ GeV. In
the most sensitive signal region (L110), at least two anti-$k_T$ jets of
distance parameter 0.4 were required to have $p_T>100,\ 50$ GeV, respectively.
Cuts on the stranverse mass variable~\cite{Lester:1999tx} $m_{T2}>110$ GeV,
the azimuthal 
angle between the jets and the $\slash \!\!\! p_T$ vector, $\Delta \phi_j>1.0$,
and the azimuthal angle between the $\slash \!\!\! p_T$ and 
$p_T=\slash \!\!\! p_T + p_T(l_1) + p_T(l_2)$, $\Delta \phi_l<1.5$, were also
employed in order to increase the expected sensitivity over
backgrounds.

\begin{figure}[t]
   \includegraphics[width=0.5\textwidth]{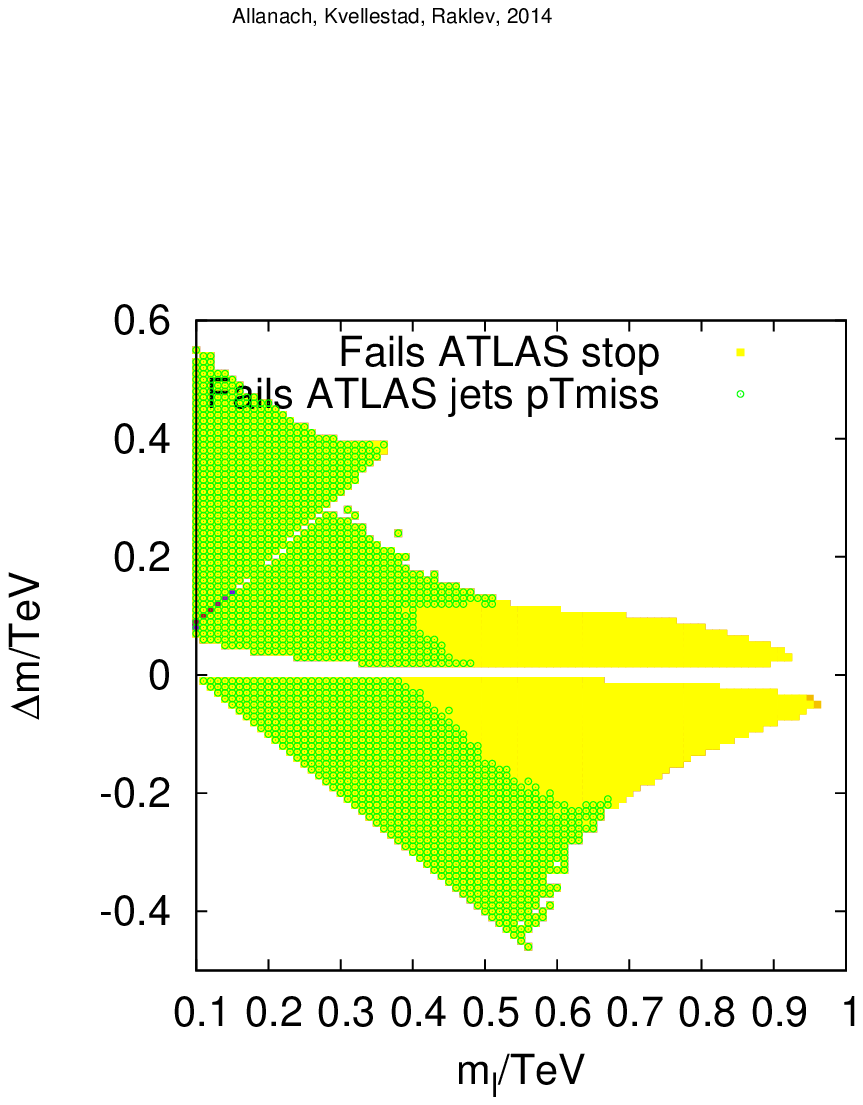}
   \caption{Constraints on the golden channel parameter space from the ATLAS 
     stop pair production search involving two leptons.
     Yellow points are ruled out by the ATLAS search for leptonically decaying
     stops at the
     95$\%$ CL, whereas those coloured green are ruled out by ATLAS
     jets plus $\slash \!\!\! p_T$ searches at the $95\%$
     CL.} 
   \label{fig:stop}
 \end{figure}

ATLAS observe 3 events on a background of $5.2\pm 2.2$, which they calculate
corresponds to a 95$\%$ CL upper bound on a putative beyond the Standard Model
contribution of 5.6 events. We show the effect on our parameter space in Fig.~\ref{fig:stop}. Seven points are left after applying this
constraint, each of which is excluded by the ATLAS jets plus $\slash \!\!\!
p_T$ search. 

As a check of the robustness  of this result we have varied other parameters than the ones shown as axes in our figures one by one: we have increased the gluino mass to 2 TeV, we have changed the slepton mass ratio $m_{{\tilde
  l}_L}(M_{SUSY})/m_{{\tilde l}_R}(M_{SUSY})=1.5,\ 2.5$, completely decoupled the left-handed slepton $m_{{\tilde l}_L}(M_{SUSY})=1$ TeV, split the squark masses $m_{{\tilde q}_R}(M_{SUSY})=m_{\tilde q}(M_{SUSY})+200$ GeV
and $m_{{\tilde q}_L}(M_{SUSY})=m_{\tilde q}(M_{SUSY})+200$ GeV. However, the
results are very similar to those shown above and the 
conclusion is identical in each case: the search for jets plus $\slash \!\!\!
p_T$ and the stop pair searche rule out the whole of the parameter space that
is consistent at 95$\%$ CL with the CMS $jjll\slash \!\!\! p_T$ excess. 

\section{Conclusions}
\label{sec:conclusion}
To summarize, we have shown that a golden cascade interpretation of the CMS
excess in  $lljj\slash \!\!\! p_T$ events is apparently viable in its own
terms. 
A recent ATLAS search using similar cuts leaves a sizeable portion of
parameter space consistent with the $lljj\slash \!\!\! p_T$ excess at the
95$\%$ CL\@. However, the interpretation 
is in tension with other sparticle searches. 
In particular, the
combination of ATLAS searches for jets plus $\slash \!\!\! p_T$, an ATLAS
di-stop search involving two leptons and a CMS four-leptons plus $\slash \!\!\!
p_T$ search
has no overlap at the 95$\%$ CL with the CMS  excess.

\acknowledgments
This work has been partially supported by STFC grant 
ST/L000385/1. We
thank the Cambridge SUSY Working Group and T.\ Stefaniak for stimulating
discussions. Some of the CPU 
intensive parts of this work was performed on the Abel Cluster, owned by the
University of Oslo and the Norwegian metacenter for High Performance Computing
(NOTUR), and operated by the Research Computing Services group at USIT, the
University of Oslo IT-department. The computing time was given by NOTUR
allocation NN9284K, financed through the Research Council of Norway. 

\bibliography{dilep}
\end{document}